\documentclass[prb,twocolumn,superscriptaddress,showpacs] {revtex4-1}
\usepackage{amsmath,amsfonts,amssymb,mathbbol}
\usepackage{graphicx,psfrag,color}
\usepackage{dcolumn}
\usepackage{bm}
\usepackage{mathbbol, appendix}

\begin{document}

\title{Viewpoint on the ``Theory of the superglass phase" and a proof of principe of quantum critical jamming and related phases}
\author{Zohar Nussinov}
\affiliation{Department of Physics, Washington University, St.
Louis, MO 63160, USA}

\date{\today}

\begin{abstract}
A viewpoint article on the very interesting work of Biroli,
Chamon, and Zamponi on superglasses. I further suggest
how additional new superglass and "spin-superglass" phases of matter
(the latter phases contain quenched disorder) and general characteristics
may be proven as a theoretical proof of concept in various electronic
systems. The new phases include: (1) superglasses of Cooper pairs, i.e.,
glassy superconductors,  (2) superglass phases of quantum spins,
and (3) superglasses of the electronic orbitals. New general features which may be derived by the same construct
include (a) quantum dynamical heterogeneities- a low temperature
quantum analogue of dynamical heterogeneities known to exist
in classical glasses and spin-glasses wherein the local dynamics
and temporal correlations are spatially non-uniform. I also
discuss on a new class of quantum critical systems.
In particular, I  outline (b) the derivation of the quantum
analogue  of the zero temperature jamming transition that
has a non-trivial dynamical exponent. We very briefly comment
on (c) quantum liquid crystals.
\end{abstract}
\maketitle

\section{Introduction}

In an exciting article in this issue, Biroli, Chamon,
and Zamponi, (BCZ) illustrate theoretically the possibility of
a ``superglass" phase \cite{BCZ}.  This new quantum phase forms an intriguing
amorphous counterpart to the "supersolid" phase \cite{supersolid, Chester}
that has seen a surge of interest in recent years \cite{KC}.
Within a "supersolid" phase, superfluidity can occur without
disrupting crystalline order.

So, what are ``superglasses"? Glasses are liquids
that have ceased to flow on experimentally measurable
time scales. By constrast, superfluids flow without any
resistance. The existence of a phase characterized by simultaneous glassiness
and superfluidity may seem like a clear contradiction
of terms. In their article, BCZ prove that
this is not so. Interacting quantum particles can indeed form such a
``super-glass" phase at very low temperature and high density; their work
confirms the earlier numerical suggestion of such a phase by Bonnsegi, Prokof'ev, and
Svistunov \cite{sg} and an investigation by Philips and Wu \cite{pw}. The superglass
phase is characterized by an amorphous density profile.  
At the same time, a finite fraction of the particles flow without any
resistance- as if they were superfluid.  Thus, the "superglass" constitutes a glassy
counterpart to the "supersolid" phase .

The approach invoked by BCZ to
prove the existence of superglasses is particularly elegant. It relies on the 
mapping \cite{map} between
viscous classical  systems whose properties are well known 
to new many body quantum systems. 
In realizing the link between classical and 
quantum systems to gain insight into the quantum many body phases, 
BCZ nicely add an important new result to earlier investigations that built 
on such similar insights elsewhere.  Chester \cite{Chester} suggested the 
existence of a supersolid by relying on such a connection. In a similar fashion, Laughlin invoked
a highly inspirational analogy \cite{Laughlin} between variational (Jastrow type) wavefunctions describing the fractional
quantum Hall system and a well known system of classical charged particles 
interacting via a logarithmic potential.  By using the classical plasma analogy and 
using known results on it, Laughlin was 
able to make headway on the challenging 
many body quantum problem and construct his highly successful
wavefunctions. The mapping used by BCZ similarly enables
exact results on quantum problem 
and a detailed correspondence of spatial and temporal correlations
between the classical and quantum systems.
BCZ apply this mapping to a classical system well known to exhibit 
glassy dynamics-  the Brownian hard sphere problem.
The quantum counterpart of the classical hard sphere
problem is a natural system containing hard sphere interactions.  On the classical side of the correspondence,
the hard sphere system
has been heavily investigated. \cite{sphere, sphereglass1, sphereglass2}
When the sphere packing density is slowly varied,  the classical Brownian hard sphere
system undergoes a transition from a liquid at low density to an ordered 
crystal at high density \cite{sphere}. When crystallization is thwarted by 
a rapid increase of the packing density or by, e.g., a change of the particle geometry,  
the system cannot order nicely into a crystal and instead jams into a dense amorphous 
glass \cite{sphereglass1, sphereglass2}.  BCZ noticed that when fused with the mapping between classical and quantum systems, information on classical glass forming systems
such as the Brownian spheres gives rise to highly non-trivial results.  In particular, the glassy phase
of the classical system translates into a quantum glass of a Bose system. Similarly, the classical solid 
maps onto a quantum bosonic crystal. The ensuing phase diagram
is provided in Fig.(2) of their article.  The spatio-temporal correlations of the (bosonic) quantum dual 
can be computed  by mapping to the classical system. Both the glassy and solid phases 
harbor a finite Bose-Einstein condensate fraction. Putting all of the pieces together, BCZ provide an important 
proof of concept of the superglass phase in a simple
and precise way. This route may be replicated for classical systems
other than the Brownian hard sphere which also display solid
and glass phases. 

What physical systems may realize the new superglass phase?
Recent experiments \cite{KC} on solid
Helium 4 exhibit super-solid  type features and have led to a flurry of activity. 
In the simplest explanation of observations, a fraction of the medium becomes,
at low temperatures, a superfluid that decouples from the measurement apparatus. 
However, the required condensate
fraction does not simply conform with thermodynamic measurement \cite{bgnt}. 
Rittner and Reppy \cite{Reppy} further found that the putative 
super-solid type feature is accutely sensitive 
to the quench rate for solidifying the liquid. Aoki, Keiderling, and Kojima 
discovered rich hysteresis and memory effects \cite{Kojima}.   All of these features 
 can arise from glassy 
characteristics alone \cite{bgnt, dorsey}- precisely as in the superglass
phase discussed by BCZ. It may indeed well be that a confluence of both superfluid 
and glassy features (and their effects on elastic properties, 
e.g., screened finite elastic shear penetration depths)  \cite{znm}  may be at work. 
There may be new experimental consequences
of (super-) glassy dynamics such as this. For instance, such dynamics
can manifest disparate relaxation times that may be probed for. \cite{bgnt} 
Typical glass formers indeed typically exhibit relaxations on two different time scales.
 
Cold atom systems may provide another realization
of a superglass state. Indeed, a supersolid state of cold atoms in a confining optical lattice 
was very recently achieved \cite{coldatom}. It is 
natural to expect a superglass analog of
these cold atomic systems. 

The super-glass phases that BCZ find and the mapping they employ may
also have new manifestations elsewhere. We suggest a few of these below. 

We may envision lattice extensions
of the continuum system investigated by BCZ:
a "lattice superglass".  For charged bosons (e.g., Cooper pairs) on a lattice,
such a {\em charge superglass} would correspond to a superconductor
with glassy dynamics. In a similar vein,
a "lattice supersolid" of Cooper pairs would correspond to a superconductor concomitant with
well defined crystalline (i.e., charge density wave) order.  
Indeed, in some heavy fermion compounds as well as in the cuprate and
the newly discovered iron arsenide family of high temperature superconductors \cite{jeh}
there are some indications of non-uniform meso-scale spatial electronic 
structures and glassy dynamics. Classical glass formers are known to 
exhibit "dynamical heterogeneities"- a non-uniform distribution of local velocities \cite{het.}.
Using the very same mapping used by BCZ-   {\em Quantum dynamical
heterogeneities} might be suggested in their corresponding quantum counterparts.
Such heterogeneities with concurrent non-uniform spatial structures 
might have realizations in some of the aforementioned electronic systems. 

Other realizations may be manifest as {\em spin superglasses}. 
Quantum spin systems in a magnetic field, \cite{kodma} can exhibit a delicate interplay
between the formation of singlet states and the tendency of spins to align with the field
direction. These systems can be mapped onto a system of bosons with
hard core interactions- just as in the system investigated by BCZ.
In some spin S=1/2 antiferromagnets in an external magnetic field, 
triplet states with spins aligned along the field direction
can be regarded as hard core bosons.  
In many other systems, interactions between quantum
spins may also be mapped onto hard core type
bosonic systems. \cite{mat}  \cite{cristian}
If a solid or glassy phase appears
in a classical Brownian system, 
then a mapping similar to that of BCZ suggests 
supersolidity/superglassiness in the corresponding quantum spin system.
Recently, there has been much work examining
supersolidity in such spin systems, e.g.,  \cite{cristian}.
It is highly natural to expect new lattice spin superglass
counterparts. 

In transition-metal compounds, the fractional filling of the 3d-shells allows for cooperative orbital ordering.  This order has been observed in numerous compounds.\cite{orbit}
Similar to the spin and charge degrees of freedom, we may ask
whether low temperature Bose condensed glasses of orbitals may appear: 
an {\em orbital superglass}. The work
of BCZ allows to investigate this by knowing the 
dynamics of hard core Bose model derived from
a classical counterpart. Orbital states can be described in terms of a S=1/2 pseudo-spin. \cite{orbit}
We may, in turn, map these pseudo-spins to
hard core bosons \cite{mat} and then investigate the dynamics of these bosons by 
mapping the system to that of classical Brownian particles on a lattice.

The mapping employed by BCZ also suggests a new
quantum critical point in related systems. 
The classical jamming transition \cite{J}  of hard spheres/disks
from a jammed system at high density to an unjammed one at lower densities
is a continuous transition with known critical exponents, both static
\cite{J1} and dynamic \cite{hatano}.
Replicating the mapping used by BCZ \cite{detail}, we may derive an analog quantum 
system harboring a zero temperature transition
with similar critical exponents. The classical zero temperature
 critical point ("point  J")  \cite{J, J1} 
may rear its head anew in the form of {\em quantum critical jamming}
of the bosonic systems with dynamical exponents, potentially as high as $z=4.6$, as we may ascertain from
those reported for the classical jamming system \cite{hatano}.

All of the above mentioned examples are free of quenched disorder. 
Glassiness in structural glasses and the Brownian hard sphere system
is not triggered by disorder. 
In classical (and quantum) spin glass system, sluggish dynamics is triggered by quenched
disorder.  \cite{parisi} Applying the same mapping of BCZ anew on viscous classical
systems with quenched disorder, we may examine
quenched {\em super spin-glass} analogs of spin-glass systems.
In a similar vein, quantum analogs of classical liquid crystalline systems
(nematic or smectic phases) first advanced in \cite{elc} and further investigated in \cite{znm} can be derived by examining the bosonic analog of classical
liquid crystalline systems. 

The work of BCZ provides important steps in an entirely new field. It proves,
as a matter of principle, the existence of a superglass phase
in a physical quantum systems. The ramifications of
such a phase may be numerous. 


\end{document}